\documentclass[aps,prb,reprint,superscriptaddress]{revtex4-2}
\usepackage{graphicx}
\usepackage{xcolor}

\begin{document}

\title{Fabrication of microstructured devices of the unconventional superconductor CeCoIn$_5$ for investigations of isolated grain boundaries}

\author{S.~Mishra} \email[]{sm335@rice.edu}
\altaffiliation{Present address: Department of Physics and Astronomy, Rice University, Houston, Texas, 77005, USA}
\affiliation{Los Alamos National Laboratory, Los Alamos, New Mexico, 87545, USA}

\author{S. M. Thomas}\affiliation{Los Alamos National Laboratory, Los Alamos, New Mexico, 87545, USA}

\author{R. McCabe}\affiliation{Los Alamos National Laboratory, Los Alamos, New Mexico, 87545, USA}


\author{E. D. Bauer}\affiliation{Los Alamos National Laboratory, Los Alamos, New Mexico, 87545, USA}

\author{F. Ronning}\affiliation{Los Alamos National Laboratory, Los Alamos, New Mexico, 87545, USA}

\date{\today}

\begin{abstract}
Grain boundaries are critical for determining the functionality of polycrystalline materials. Here we present on the structural $\&$ transport properties of grain boundaries in the unconventional superconductor CeCoIn$_5$. We provide a detailed recipe for the fabrication of isolated grain boundary devices from of as-grown polycrystalline samples of CeCoIn$_5$. Electron backscattered diffraction imaging of polycrystalline CeCoIn$_5$ samples reveals an abundance of $90^\circ$ misorientation grain boundaries suggesting a preferential nucleation of CeCoIn$_5$ grains with 90$^\circ$ misorientation over a random distribution of grain orientations. Transport measurements across grain boundary devices establish coherence of superconductivity and allows us to establish a lower bound on the critical current density for the grain boundaries. Our work opens new possibilities for fabrication of quantum devices such as Josephson-junctions out of bulk unconventional superconducting materials.
\end{abstract}

\maketitle

\section{Introduction}
Microscopic interactions dictate the phase of matter that exists in a material, but the grain boundaries and microstructure are key to determining a material's functionality. For superconducting applications, grain boundaries can act as pinning sites for vortices and can form weak links that support the Josephson effect. In unconventional superconductors, grain boundaries are critical for establishing phase sensitive measurements of the order parameter, as demonstrated on the cuprate superconductors \cite{Wollman1993, Tsuei1994, Tsuei2000}. The predominant $d_{x^2 - y^2}$ symmetry was firmly established by the observation of a half-magnetic flux quanta in a sub-micron ring patterned on a cuprate superconducting film deposited on a tri-crystal substrate.

\begin{figure*}
 \centering
  \includegraphics[width=\linewidth]{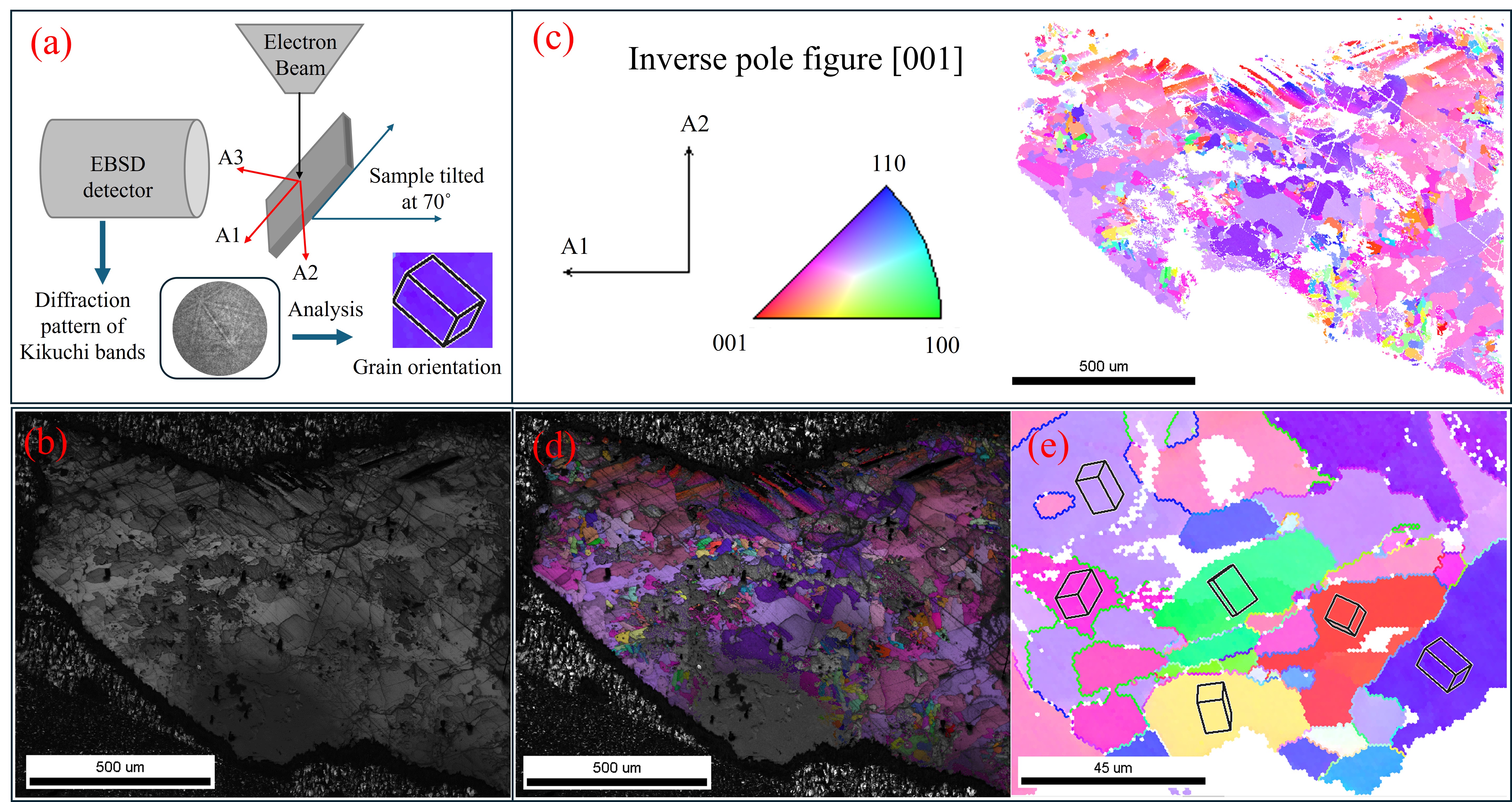}
\caption{\label{fig:EBSDtechplusIPF2} Grain and grain boundary imaging of a polycrystaline CeCoIn$_5$ sample using EBSD. (a) Schematic diagram of the EBSD imaging setup. (b) Image quality (IQ) map demonstrating the quality of collected diffraction patterns in grayscale as discussed in the text. (c) Inverse pole figure (IPF) map for the sample normal along the $[001]$ crystal lattice direction representing different grain orientations specified by the color-coded label on the left. (d) IPF map overlaid on an grayscale IQ map. (e) Enlarged view of a small region of IPF in (c) depicting the grain boundaries between neighboring grains as well as orientation of individual grains through an overlaid tetragonal unit cell.}
\end{figure*}

However, such experiments on a variety of exotic superconductors are challenging due their unavailability in thin-film form and subsequently a lack of comprehensive understanding of their grain-boundaries and weak-link behavior. Therefore, to investigate the feasibility of such experiments in bulk superconducting materials, a recipe for fabrication of isolated grain boundary devices from bulk polycrystalline samples is needed. In this regard, the unconventional superconductor CeCoIn$_5$ is an ideal candidate material. CeCoIn$_5$ is a well-studied heavy fermion superconductor below $T_c = 2.3$ K \cite{CPetrovic} and a $d_{x^2 - y^2}$ superconducting order-parameter pairing symmetry was previously determined via phase-sensitive measurements \cite{Park, zhou2013, allan}. We used a combination of mechanical thinning, advanced scanning electron microscope (SEM) imaging, energy-dispersive X-ray spectroscopy (EDS), electron backscatter diffraction (EBSD), and focused ion-beam (FIB) milling to fabricate isolated grain boundary devices and explore their transport properties for weak-link behavior. Similar approaches have been used on polycrystaline cuprate samples to generate grain boundary devices \cite{Babcock1990, Schindler1992, Schindler1994, Field1997, Saleh1998}. 

In this paper, we investigate isolated grain boundaries sampled from bulk polycrystalline CeCoIn$_5$. To understand the properties of grain boundaries, we isolate and fabricate microstructure devices featuring single grain boundaries from thinned polycrystalline samples of CeCoIn$_5$ and perform electrical transport measurements. The nucleation of grains in the polycrystalline material is such that a 90$^\circ$ misorientation is preferred over a random distribution of grain boundaries. Electrical transport across the grain boundary establishes coherence of superconductivity suggesting the possibility of making Josephson-junction based devices.

\section{Experimental details}


Polycrystalline samples of CeCoIn$_5$ were synthesized by heating stoichiometric amounts of the constituent elements in an alumina crucible sealed under vacuum within a quartz tube to 1100 $^\circ$C, and cooling to 900 $^\circ$C, and then quenching in liquid nitrogen. After, the samples were annealed at 600 $^\circ$C for 1 month and then etched in dilute HCl to remove excess free In.

The polycrystalline samples of CeCoIn$_5$ grew in the form of fused ingots. Shiny facets corresponding to single grains or microcrystals of CeCoIn$_5$ are apparent on the surface of the ingots. For grain imaging and micro-machining, the samples are polished down to plates (thickness $\sim$ 10-20 $\mu$m) with ultra-flat faces of roughness down to $\sim$0.04~$\mu$m. The details of sample preparation for EBSD imaging is outlined in Appendix A. The resulting sample surfaces allowed clean imaging of CeCoIn$_5$ grains.

Energy-dispersive X-ray spectroscopy performed on the polished sample surface reveals that along with CeCoIn$_5$, the sample contains unwanted impurity phases, namely the cubic antiferromagnet CeIn$_3$ \cite{Knebel} and the tetragonal heavy fermion superconductor Ce$_2$CoIn$_8$ \cite{YAMASHITA}. In addition, a few small patches of free cobalt (mostly $<20~\mu m^2$) are also observed to be sparsely scattered on the sample surface. Nonetheless, in all samples, CeCoIn$_5$ is the dominant phase with a volume fraction greater than 60 - 70$\%$ compared to CeIn$_3$ (10-15$\%$) and Ce$_2$CoIn$_8$ (10-15$\%$). The EDS results are shown in Appendix B.


\begin{figure*}
	\centering
   \includegraphics[width=\linewidth]{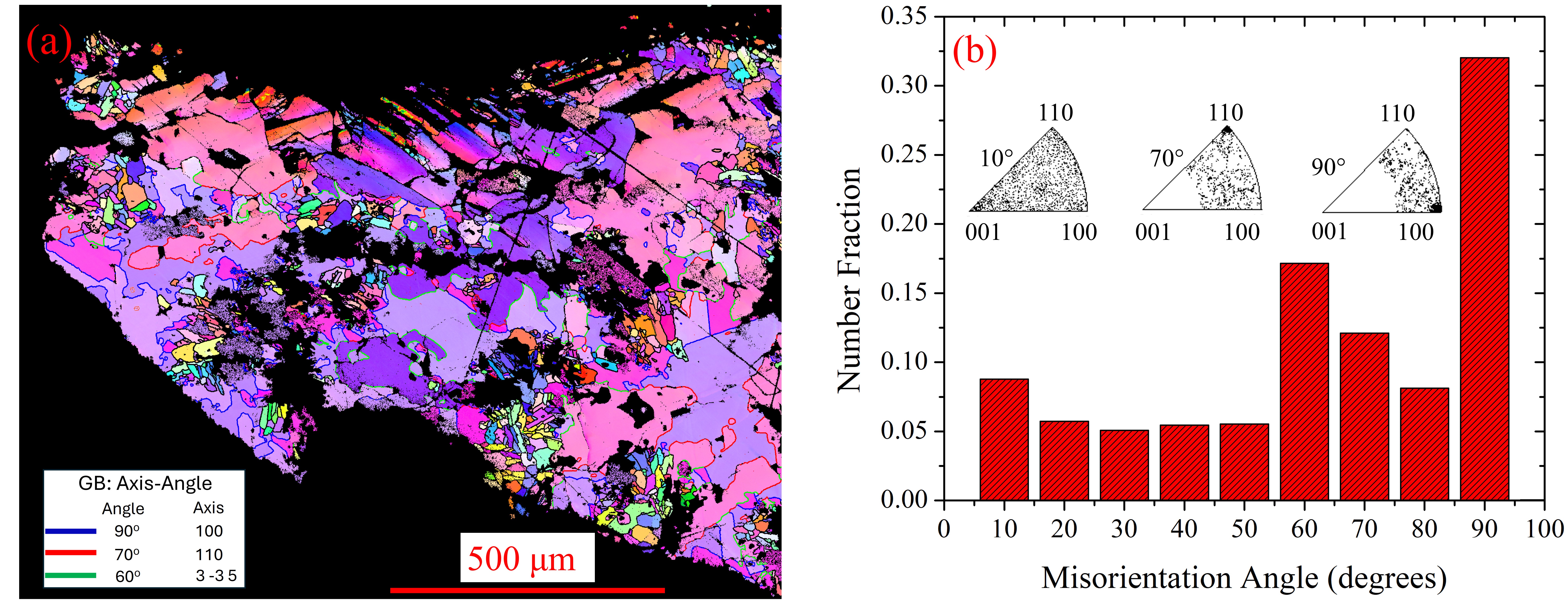}
\caption{\label{fig:misorientation2} Identification of grain boundaries between neighboring grains in a polycrystalline sample of CeCoIn$_5$. (a) Inverse pole figure depicting grain boundaries between the neighboring grains misoriented by angles 90$^\circ$ about $[100]$ tetragonal direction (solid blue line), 70$^\circ$ about $[110]$ direction (solid red line) and 60$^\circ$ about $[3\bar{3}5]$ direction (solid green line), respectively. (b) Histogram illustrating the fractional distribution of grain boundaries with misorientation angles between 0$^\circ$ to 90$^\circ$, binned in interval size of 10$^\circ$. Inset shows the fractional distribution of grain boundaries projected on an inverse pole figure for the misorientation angles 10$^\circ$, 70$^\circ$ and 90$^\circ$.}
\end{figure*}

\section{CeCoIn$_5$ grain imaging with EBSD}
The structural and crystallographic imaging of the grains and grain boundaries present in the CeCoIn$_5$ samples is performed using electron backscatter diffraction imaging. EBSD is a surface-sensitive technique and requires high-quality surfaces for indexing. EBSD was performed on a Thermo Fisher Scientific Apreo SEM using a EDAX Velocity EBSD camera. A schematic of EBSD setup is shown in Figure \ref{fig:EBSDtechplusIPF2}(a) and the corresponding details are provided in Appendix C. The collected EBSD patterns were indexed to the CeCoIn$_5$ crystal structure \cite{Kalychak} to obtain the orientation of the different grains present in the sample.

Figures \ref{fig:EBSDtechplusIPF2} (b)-(e) show the grain and grain boundary images obtained for the polished sample of polycrystalline CeCoIn$_5$. Here, the diffraction patterns are collected by scanning the electron beam (accelerating voltage 20 kV and beam current 13 nA) over the sample surface in square grids of area resolution of 1 $\mu$m $\times$ 1 $\mu $m. An EBSD image quality (IQ) map is shown in  Fig.1 (b). In this grayscale IQ map, lighter and darker color represent regions with higher and lower image quality, respectively. Based on correlative EDS analysis (shown in Appendix B), the grains with lower IQ tend to  be from CeIn$_3$ and the dark spots correspond to free Co. The Ce$_2$CoIn$_8$ impurity phase cannot be distinguished from CeCoIn$_5$ on the grayscale image quality map as similar to CeCoIn$_5$, Ce$_2$CoIn$_8$ also crystallizes in the tetragonal crystal structure \cite{Kalychak} and therefore the Ce$_2$CoIn$_8$ diffraction patterns are indexed together with CeCoIn$_5$. Spectral mapping of the elemental composition with EDS is used to distinguish the CeCoIn$_5$ phase from the Ce$_2$CoIn$_8$ impurity regions (see Appendix B).

\section{Grain misorientation and grain boundary statistics}
Figures \ref{fig:EBSDtechplusIPF2}(c)-(e) depict information about the grains and the grain boundaries present in the polycrystalline CeCoIn$_5$ sample. The grain orientations are depicted through a color-coded map of stereographic projections onto an inverse pole figure (IPF) map. In the IPF maps shown in Fig. \ref{fig:EBSDtechplusIPF2}(c)-(e), red represents the $[001]$ direction of a tetragonal CeCoIn$_5$ grain, blue represents $[110]$, and green represents $[100]$ along the sample normal direction A3. Distinctly colored grains are overlaid with a tetragonal unit cell oriented accordingly in Fig. \ref{fig:EBSDtechplusIPF2}(e). Information about grain sizes can also be inferred from the IPF map. It is evident that the CeCoIn$_5$ grain sizes vary from $\sim 5-100$ $\mu$m. When the IPF map is overlaid on the gray-scale IQ map [see Fig. \ref{fig:EBSDtechplusIPF2}(d)] it becomes obvious that the darker regions in the IQ map cannot be indexed against the CeCoIn$_5$ crystal structure. Having collected the information about orientations and sizes of the grains, the grain boundaries and grain boundary orientations between neighboring grains can be identified as shown in Fig. \ref{fig:EBSDtechplusIPF2}(e).


We quantitatively analyze these grain boundaries in terms of a conventional axis-angle misorientation representation where the grain boundaries between neighboring grains are identified in terms of their relative misorientation about a common axis. In Fig. \ref{fig:misorientation2}, we present axis-angle misorientation analyses of grain boundaries in the polycrystalline CeCoIn$_5$ sample through an inverse pole figure map [Fig. \ref{fig:misorientation2}(a)] and a quantitative histogram [Fig. \ref{fig:misorientation2}(b)]. The histogram illustrates the fractional distribution of grain boundaries with respect to the misorientation angle, binned in interval size of $10^\circ$. The insets in the histogram show the number fraction of grains for three misorientation angles 10$^\circ$, 70$^\circ$ and 90$^\circ$ about different misorientation axes in IPF representation. It is evident from the density of points representing grains that the largest fraction of grains is misoriented by 90$^\circ$ about the $[100]$ direction. This grain boundary misorientation is indicated by solid blue lines in Fig. \ref{fig:misorientation2}(a) and reflects two adjoining grains oriented such that their $c$ axes are orthogonal to each other and they share an in-plane $a$ direction. The preferential misorientation of 90$^\circ$ about the $[100]$ direction provides important insights about the crystal growth of polycrystalline CeCoIn$_5$. The parent compound CeIn$_3$ crystallizes in a cubic structure and tetragonal CeCoIn$_5$ can be obtained by alternating layers of CeIn$_3$ and CoIn$_2$ stacked along the $c$ axis. Therefore, a 90$^\circ$ misorientation about $[100]$ can be explained as the growth of two CeCoIn$_5$ grains on two orthogonal faces of a cubic CeIn$_3$ nucleation site as sketched in the cartoon in Fig. \ref{fig:Nucleationcartoon}(a). It should be noted that grain boundary information is only valid at the surface and how it evolves inside the sample cannot be determined with the surface EBSD data alone. Other misorientations such as 70$^\circ$ about $[110]$ direction are also relatively prominent in the sample. 

\section{Fabrication of grain boundary devices}
\begin{figure}
	\centering
   \includegraphics[width=\columnwidth]{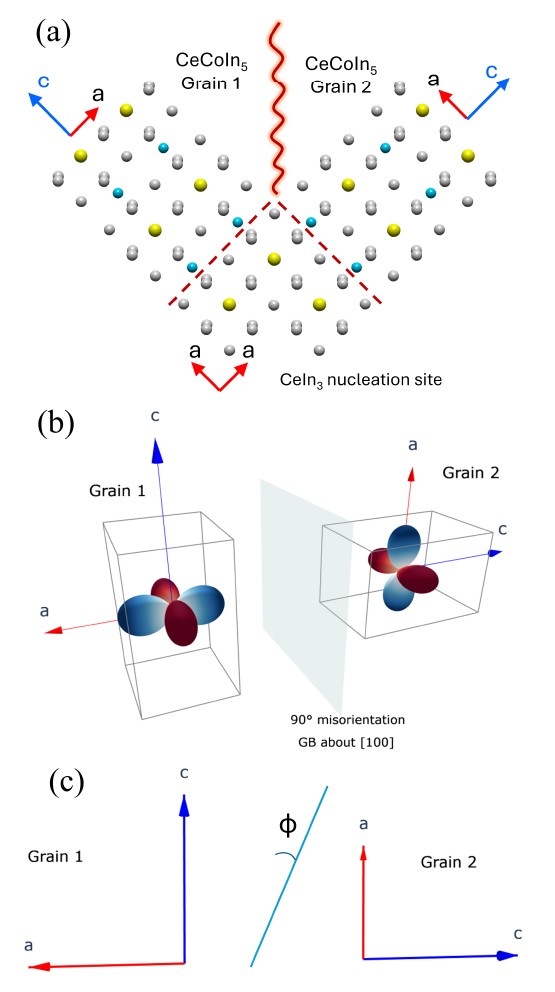}
\caption{\label{fig:Nucleationcartoon} Illustration of 90$^\circ$ grain boundary in polycrystalline CeCoIn$_5$ sample. (a) Cartoon depicting two tetragonal CeCoIn$_5$ grains growing at adjacent orthogonal faces of a CeIn$_3$ nucleation site, leading to a 90$^\circ$ misorientation grain boundary. Yellow, cyan and gray spheres are cerium, cobalt, and indium atoms, respectively. (b) $d_{x^2-y^2}$ SC order parameter overlaid on the tetragonal CeCoIn$_5$ lattice across the 90$^\circ$ grain boundary about $[100]$. (c) 90$^\circ$ grain boundary in (b) oriented at an angle $\phi$ with respect to the tetragonal $c$ axis as discussed in the text.}
\end{figure}

\begin{figure*}
	\centering
  \includegraphics[width=\linewidth]{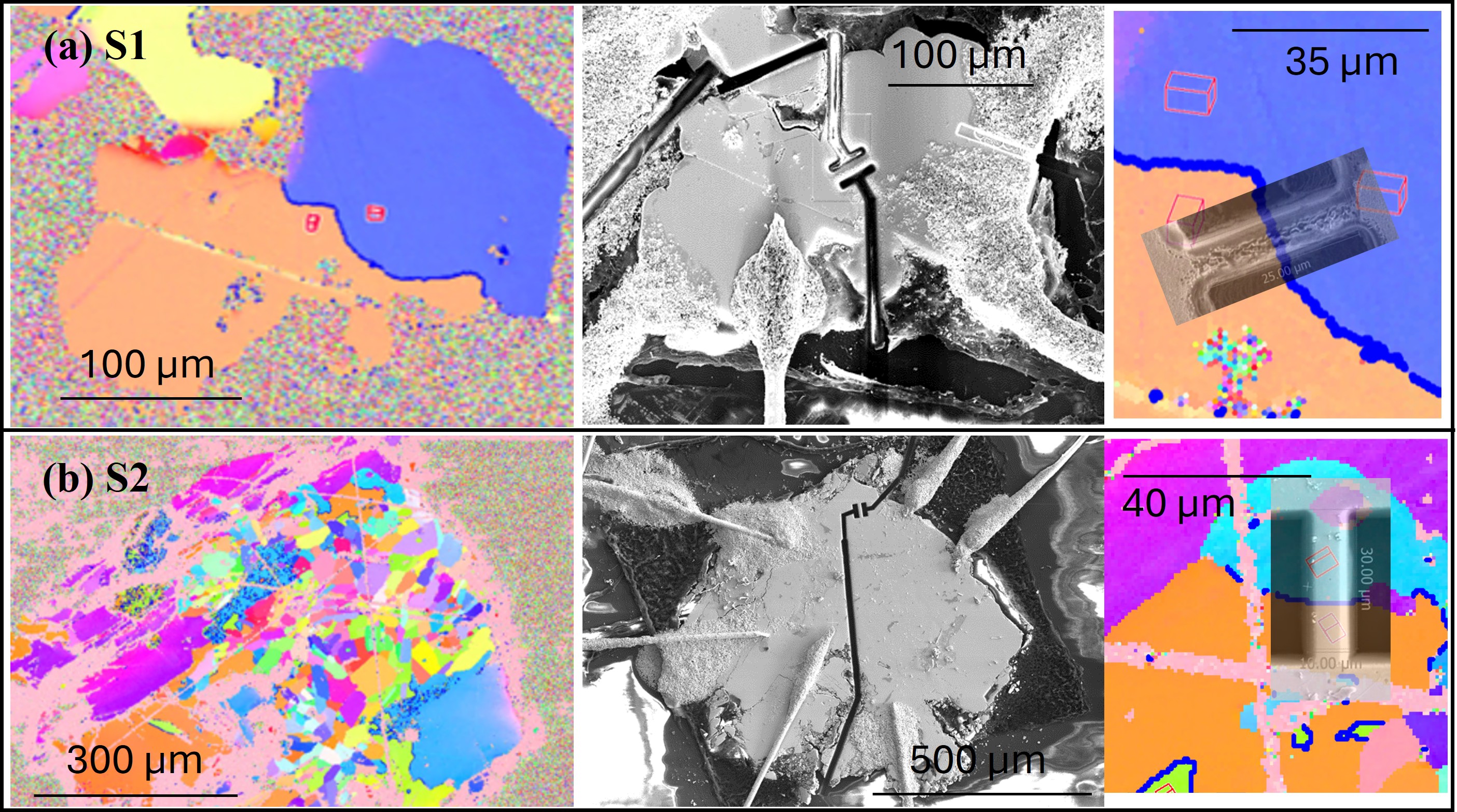}
\caption{\label{fig:Transportdevicesmain} Fabrication of microstructured devices across an isolated grain boundary for the samples (a) $S1$ and (b) $S2$. (Left panels) Inverse pole figure maps of a polycrystalline CeCoIn$_5$ sample with a 90$^\circ$ misorientation grain boundary about $[100]$ depicted by the solid blue line. (Middle panel) SEM images of the fabricated grain boundary devices. (Right panel) Enlarged view of the 90$^\circ$ misorientation grain boundaries with an overlaid SEM image of the corresponding devices.}
\end{figure*}

\begin{figure*}
	\centering
  \includegraphics[width=\linewidth]{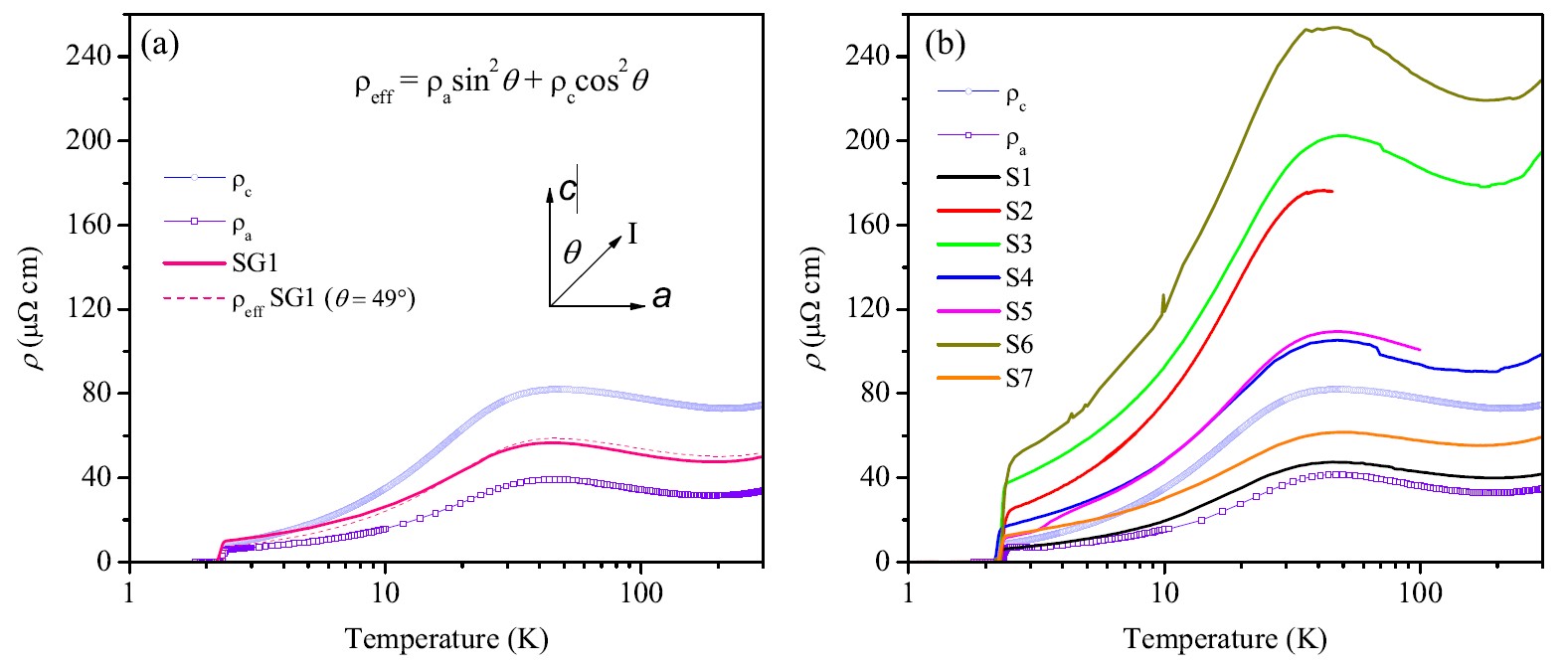}
\caption{\label{fig:RvsT} Temperature-dependent resistivity $\rho (T)$ of CeCoIn$_5$ (a) single crystal, single grain devices and (b) grain boundary devices. $\rho_c (T)$ data is taken from Ref. \cite{Malinowski}. Dashed lines in (a) corresponds to $\rho_{eff}$ as discussed in the text.}
\end{figure*}

Using the information about the grains and the grain boundaries in the samples, we fabricate isolated grain boundary devices to study the electrical properties of individual grains and/or grain boundaries. In this regard, the above discussed 90$^\circ$ misorientation grain boundaries about $[100]$ are particularly interesting not only because of their abundance but also due to their superconducting properties. In an unconventional superconductor the critical current can be completely suppressed by the correct misorientation of two grains relative to a grain boundary \cite{Sigrist1992} as has been observed in cuprate superconductors \cite{Mannhart, Eom1991} and illustrated through Fig. \ref{fig:Nucleationcartoon}(b). For a 90$^\circ$ grain boundary about [100] the critical current is proportional to sin(2$\phi$) where $\phi$ is the angle between the $c$ axis and the grain boundary [see Fig. \ref{fig:Nucleationcartoon}(c)]. Thus, if $\phi =$ 0 or 90$^\circ$ there will be a vanishingly small $J_c$. Therefore, a majority of the devices we fabricated feature a 90$^\circ$ misorientation grain boundary in an attempt to witness a suppressed critical current due to the superconducting order parameter.

After EBSD imaging, the sample was prepared for micromaching according to the details listed in Appendix D. The sample with electrical contacts is moved to an SEM equipped with focused ion beam (FIB) milling capability. A one-to-one correspondence between the sample images obtained from EBSD, SEM and optical microscope was established to identify and isolate the desired grain boundary on the sample. Using FIB, rectangular bridge-like devices (width 5-10 $\mu$m $\&$ length 20-30 $\mu$m) were fabricated across the desired grain boundary as shown in Fig. \ref{fig:Transportdevicesmain}  for two representative samples $S1$ and $S2$. The corresponding IPF map depicting the 90$^\circ$ misorientation grain boundary through a solid blue line overlaid with the SEM image of the device fabricated across the boundary is shown in the right panels of Fig. \ref{fig:Transportdevicesmain}(a) and (b). Similarly, several other grain boundary and single grain devices were fabricated as shown in Fig. \ref{fig:Device1} and Fig. \ref{fig:Device2} in Appendix E. In total, 8 grain boundary devices and 1 single grain device were successfully fabricated and measured. 


\section{Electrical transport in grain boundary devices}

The electrical resistivity measurements of the bridge-shaped devices were performed in a standard four-point configuration using an ac-resistance bridge (Lakeshore model 372) in a Quantum Design physical property measurement system (QDPPMS). The temperature-dependent resistivity $\rho(T)$ for several samples including flux-grown single crystal, single grains and grain boundary devices fabricated out of polycrystalline samples is shown in Fig. \ref{fig:RvsT}. The electrical resistivity of CeCoIn$_5$ is anisotropic with room temperature values of $\rho_{a} \sim 35~ \mu \Omega$cm and $\rho_{c} \sim 75~ \mu \Omega$cm \cite{Malinowski}. 
It is evident from Fig. \ref{fig:RvsT}(a) that the magnitude of $\rho(T)$ of the devices fabricated on single grain [see Fig. \ref{fig:Device2} in Appendix E] is between $\rho_{a}(T)$ and $\rho_{c}(T)$, as expected for an arbitrarily oriented single grain of CeCoIn$_5$. Using the orientation of the single grains determined through EBSD, the effective resistivity $\rho_{eff}$ is calculated for the device as $\rho_{eff}=\rho_a~sin^2\theta+\rho_c~cos^2\theta$, where $\theta$ is the angle between the current direction and the $c$ axis of the single grain. The calculated $\rho_{eff}$ is shown by dashed line in Fig. \ref{fig:RvsT}(a). The only noticeable difference is that the residual resistivity of the arc melted grown grains is larger than that of the flux grown single crystals. $T_c$, however, remains the same.

$\rho(T)$ for seven grain boundary devices is shown in Fig. \ref{fig:RvsT}(b). All of the grain boundary devices show a zero-resistive superconducting state below 2.3~K as found in single crystal and single grain devices.  For a majority of the grain boundary devices, $\rho(T)$ is consistently higher than the intrinsic resistivities $\rho_{a}(T)$ and $\rho_{c}(T)$ naively suggesting additional scattering contribution from the grain boundary providing a weak link between the two superconducting grains below $T_c$. Thus, it is surprising that $T_c$ of the weak link is not suppressed relative to the individual grains.

\begin{figure}
	\centering
  \includegraphics[width=\columnwidth]{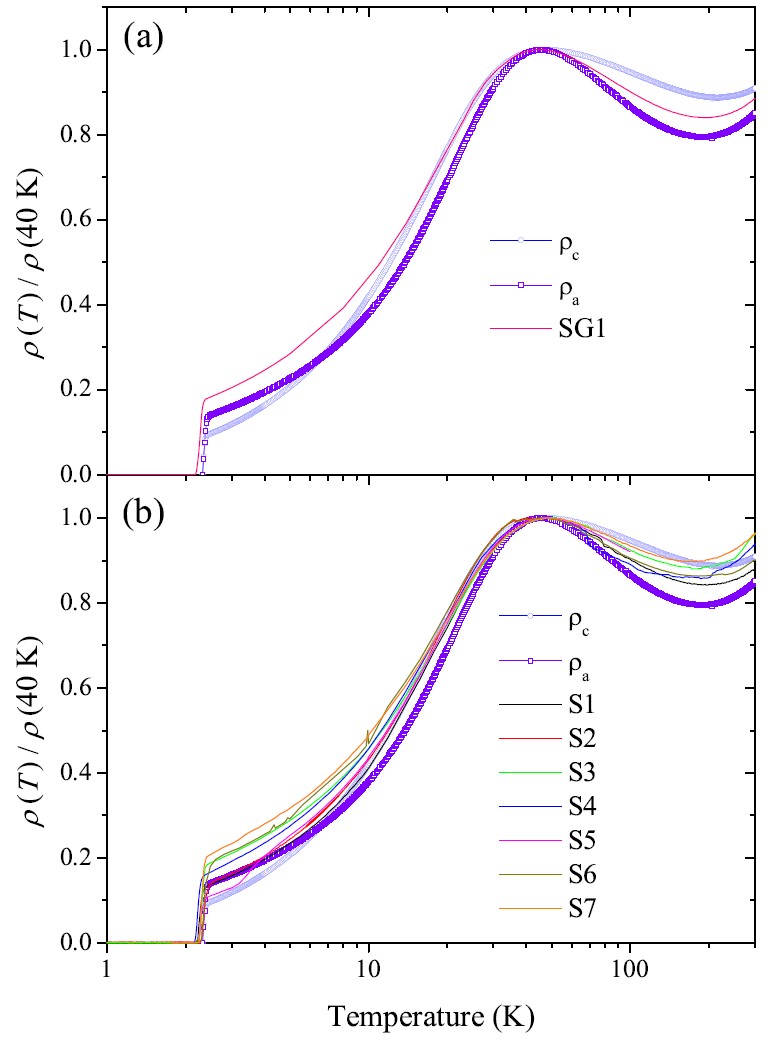}
\caption{\label{fig:RvsTnorm} Temperature-dependent resistivity normalized by the resistivity maxima at $T_{coh}=40~\text{K}$ for CeCoIn$_5$ (a) single crystal, single grain and (b) grain boundary devices.}
\end{figure}

\begin{figure*}
	\centering
  \includegraphics[width=\linewidth]{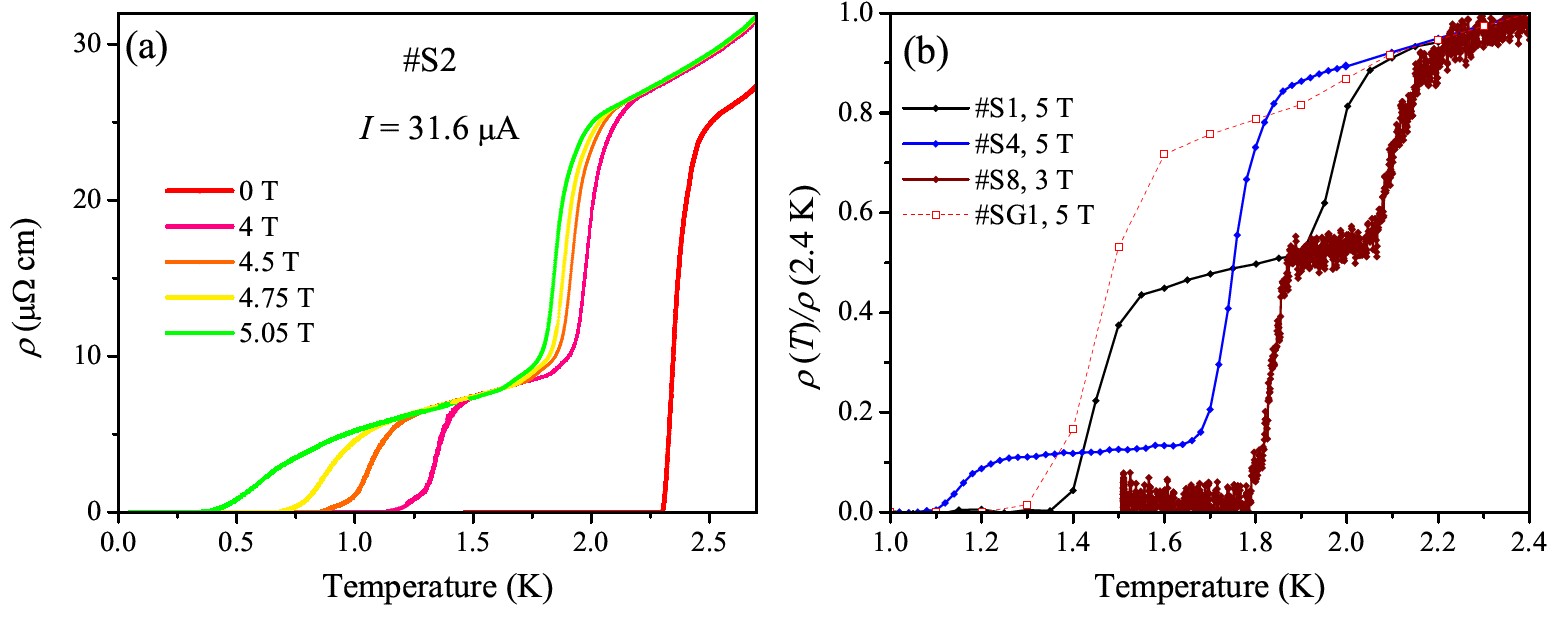}
\caption{\label{fig:RvsTvsB} Temperature-dependent resistivity at several magnetic fields for the grain boundary devices (a) $S2$ and (b) $S1$, $S4$, and $S8$ depicting distinct superconducting transitions for the differently oriented grains on either side of the grain boundary. Open symbols with dashed lines in (b) represent $\rho(T)$ for the single grain device $SG1$ featuring a single superconducting transition.}
\end{figure*}

To investigate this apparent contradiction in Fig. \ref{fig:RvsTnorm} we scale $\rho(T)$ for all devices by the resistivity maxima at $T~=~40$ K. The scaling is nearly identical for single grain and grain boundary devices shown in Fig. \ref{fig:RvsTnorm}(a) and \ref{fig:RvsTnorm}(b), respectively. To explain this we envision the following two possibilities for the grain boundary resistance $R_{GB}$. Either the grain boundary resistance is negligible compared to the resistance of the individual grains and the geometric factor is incorrect, or the grain boundary resistance scales with pure CeCoIn$_5$ i.e., $\rho_{GB} \propto \rho_{a}$. The latter is unlikely given that a grain boundary is a structural irregularity separating periodic regular lattices of differently oriented grains. Irrespective of the details of the atomic arrangement and type of dislocations in the grain boundary, $R_{GB}$ is not expected to have the same temperature-dependence as bulk CeCoIn$_5$. The geometric factor ($\frac{A}{l}$), where $A$ is the device cross-section and $l$ is the length, is measured precisely using SEM for all the devices. Thus, for the geometric factor to be incorrect, the current path would have to deviate from its expected path through the rectangular bridge-shaped geometry.

\begin{figure*}
	\centering
  \includegraphics[width=\linewidth]{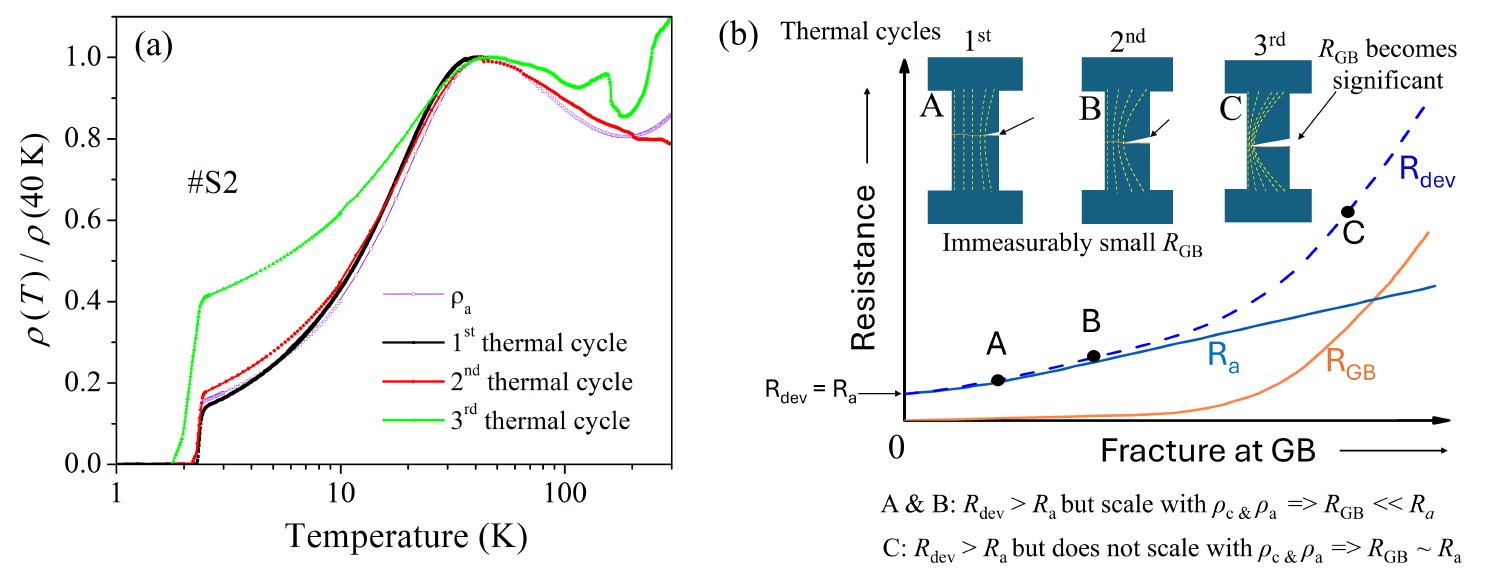}
\caption{\label{fig:B2S2andDevice} Effect of multiple thermal cycles on a fragile grain boundary device. (a) Temperature-dependent resistivity of the 90$^\circ$ misorientation grain boundary device $S2$ normalized with its maximum value at $T_{coh}=40$ K upon multiple thermal cycles. (b) Cartoon diagram and schematic plot depicting the fragility of grain boundary devices and the corresponding increase in device resistance R$_{dev}$ when put through multiple thermal cycles. Upon successive thermal cycles, the cross-section at the grain boundary reduces (with increasing fracture at GB) until the connection at grain boundary ultimately breaks.}
\end{figure*}

To illustrate that the current is indeed flowing through the device across the grain boundary we show the magnetic field dependence of the superconducting transition in Fig. \ref{fig:RvsTvsB}. The superconducting upper critical field for CeCoIn$_5$ is anisotropic with $\mu_0H_{c2} \parallel c = 5$ T and $\mu_0H_{c2} \parallel a = 12$ T \cite{Kumagai}. Therefore, when a magnetic field is applied to a grain boundary device, the resistive transition should feature two distinct superconducting drops, one for each grain with a distinct crystallographic orientation with respect to the applied field direction. Figure \ref{fig:RvsTvsB}(a) shows the superconducting transition in grain boundary device $S2$ at a few different magnetic fields showing two distinct drops in the superconducting transition attesting that the electrical transport occurs across the grain boundary in the bridge-shaped device. Similar observations in other representative grain boundary devices are shown in Fig \ref{fig:RvsTvsB}(b). The resistance drop at the higher transition temperature corresponds to the grain with its $a$ axis more closely aligned with the field direction, while the drop at the lower transition temperature corresponds to the grain with its $c$ axis more closely aligned with the applied magnetic field direction. In contrast, as expected for a single grain device, only a single drop persists in applied field as shown for the single grain device $SG1$ at 5 T in open square in Fig \ref{fig:RvsTvsB}(b).


Having established that transport definitely occurs across the grain boundary, we are left to hypothesize that the resistance across the grain boundary is not uniform. This hypothesis is supported by the fragility of our grain boundary devices. The samples are very thin $\sim 10-20~\mu$m and are held together by a single grain boundary. Many samples crumbled into pieces upon removal from the polishing fixture when dissolving the crystal bond in acetone (process outlined in Appendix D) suggesting the grain boundaries in the sample are structurally weak. Other samples remained visually intact after both the removal $\&$ micromachining process, but did not survive cooling down in the cryostat. Despite showing reasonable resistivities at room temperature, the differential thermal contraction of the sample and the sapphire substrate appears to provide sufficient strain to break many of our grain boundary devices. The grain boundary devices we present here are the ones that survive at least one cool down to the superconducting state.

We thus conclude that the grain boundary resistances $R_{GB}$ in Fig. \ref{fig:RvsT} and \ref{fig:RvsTnorm} are negligibly small compared to the total resistance of the device $R_{dev}$ given by $R_{dev}$ = $R_a$ + $R_{GB}$, where $R_a$ is the resistance contribution from the grains proportional to intrinsic CeCoIn$_5$ resistivity $\rho_a$ or $\rho_c$. Fortunately, the same fragility that destroyed many devices enabled a systematic study of grain boundary device $S2$ over multiple thermal cycles while observing its temperature-dependent resistivity. $R(T)$ of the device gradually increased upon successive thermal cycles and ultimately `breaks' after the third cycle. The resistance at 10 K gradually increased from A to B to C [see Fig. \ref{fig:B2S2andDevice}(b)] from the $1^{st}$ to the $3^{rd}$ thermal cycle of device $S2$. Notably, $R(T)$ normalized by the resistance at $T = 40$ K scales reasonably well between the $1^{st}$ and $2^{nd}$ thermal cycles but deviates for the $3^{rd}$ thermal cycle [see Figure \ref{fig:B2S2andDevice}]. Between the $1^{st}$ and $2^{nd}$ thermal cycles the increase of $R(T)$ and scaling of $R(T)/R(40 K)$ suggest that although the cross-section of the grain boundary decreases, the contribution of the grain boundary resistance $R_{GB}$ to the measured device resistance is still immeasurably small i.e. $R_{GB}$ $\ll$ $R_a$, $R_{dev}$. Thus, the enhanced $R(T)$ must be accounted for by contribution from the grains $R_a$ resulting from the funneling of current through smaller and smaller areas as the GB cross-section shrinks with thermal cycles.  For the $3^{rd}$ thermal cycle, however, $R(T)$ no longer scales reflecting a noticeable contribution of $R_{GB}$ to the overall resistance of the device. The above interpretation of grain boundary fragility is pictorially depicted along with a qualitative schematic plot in Fig \ref{fig:B2S2andDevice}(b) where upon successive thermal cycles the grain boundary cross-section gradually decreases leading to enhanced device resistance and subsequently a measurable contribution from $R_{GB}$.

\begin{figure}
	\centering
  \includegraphics[width=\columnwidth]{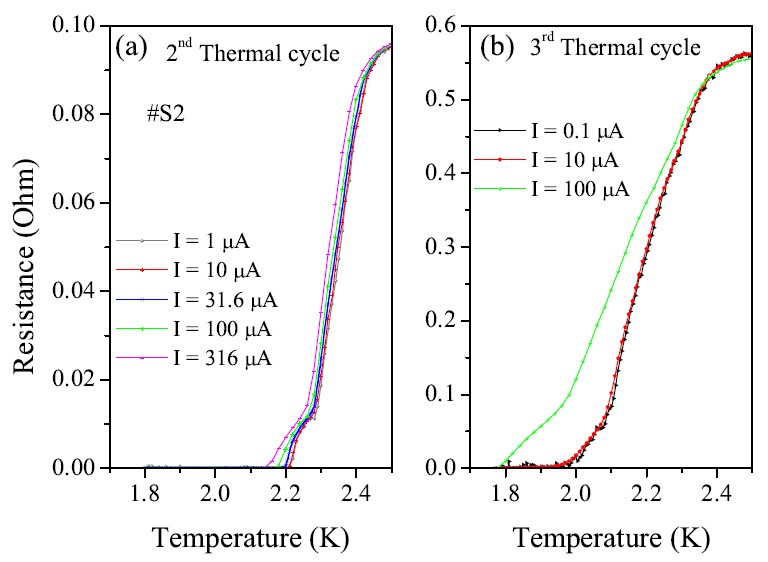}
\caption{\label{fig:B2S2RvsCurrent} $R(T)$ for the 90$^\circ$ GB device $S2$ at several different constant excitation currents during the (a) $2^{nd}$ and (b) $3^{rd}$ thermal cycles.}
\end{figure}

The evolution of $R(T)$ in device $S2$ over successive thermal cycles provides important insights regarding the coherence of superconductivity or Josephson tunneling of Cooper pairs across the grain boundary. It is important to note that even though $R(T)$ increases upon successive thermal cycles, a zero-resistance state is still realized below the superconducting transition temperature, as shown in Fig. \ref{fig:B2S2andDevice}(a). Additionally, the temperature at which the zero resistance state is achieved moves to progressively lower temperatures as is expected of weak link behavior. To get an estimate of the critical current, we performed $R(T)$ measurements at several different constant current excitations for the $2^{nd}$ and $3^{rd}$ thermal cycles of $S2$ as shown in Fig. \ref{fig:B2S2RvsCurrent}. In Fig. \ref{fig:B2S2RvsCurrent}(b) one can see that the critical current at 1.8 K is 100 $\mu$A. From the Ambegaokar Baratoff relation \cite{Ambegaokar1963} one expects the $T = 0$ critical current to be $I_c = \frac{\pi\Delta}{2eR_N}$ = 1.4 mA, where $\Delta = 535 \mu$V is the superconducting gap \cite{zhou2013} and $R_N$ is the junction resistance. Close to $T_c$ one expects the critical current to be linearly suppressed to zero providing reasonable agreement with the value of 100 $\mu$A we find at 1.8 K. This establishes that the grain boundaries in CeCoIn$_5$ support phase coherent superconductivity across them. The significant grain-boundary resistance indicates that a weak-link or Josephson behavior can be supported by as-grown boundaries in CeCoIn$_5$ provided the fragile grain boundary fabrication and transport measurements are well-controlled. Such control may be achieved with more evolved FIB fabrication techniques, such as a cross-sectional thinning, lamella liftoff, and strain-free suspension of grain boundary devices \cite{Guo}.

\section{Summary}
In conclusion, we present on the fabrication and structural $\&$ transport properties of grain boundaries in polycrystalline samples of the unconventional superconductor CeCoIn$_5$. We present a detailed recipe for the fabrication of isolated grain boundary devices from as-grown polycrystalline samples of CeCoIn$_5$. Electron backscattered diffraction imaging of well-polished samples reveals the preferential growth of CeCoIn$_5$ grains with 90$^\circ$ misorientation providing critical insight into the nucleation of layered CeMIn$_5$ (M = Co, Rh, Ir) crystals. Transport measurements across grain boundary devices establish coherence of superconductivity and indicate the possibility of making Josephson-junctions based devices from these microstructures. Our work opens new possibilities for the fabrication of quantum devices from bulk materials.

\section{Appendix}
\subsection{Sample preparation for EBSD imaging}
The polycrystallline CeCoIn$_5$ ingot was fragmented into smaller pieces $<$1 mm$^3$. Using a hand-held lapping fixture, the small fragments of polycrystalline material were rough polished on two opposite faces using a 30 $\mu$m grit aluminum oxide lapping film to obtain flat plate-like samples with thickness $\sim$200 $~\mu$m. For finer polishing, the size of the grit of the aluminum oxide lapping films was sequentially reduced from 30 $\mu$m to 1 $\mu$m and a very smooth sample surface free of any large or deep scratches (no more than 2-5 $\mu$m) was achieved with the plate thickness $\sim$25$ \mu$m. For the final finishing, the sample was first carefully polished using a 0.3 $\mu$m grit lapping film to remove the small scratches by eliminating the top 4-5 $\mu$m of the material from the surface and then a very fine scratch-free, near mirror-like surface was achieved by a final polishing of the sample on a polishing cloth with colloidal silica suspensions of grain size 0.04 $\mu$m. As a result of this process, we obtained high-quality sample surface necessary for the EBSD imaging of CeCoIn$_5$ grains.

\subsection{EDS spectral imaging and correlative analysis with EBSD maps}
\begin{figure*}
	\centering
   \includegraphics[width=\linewidth]{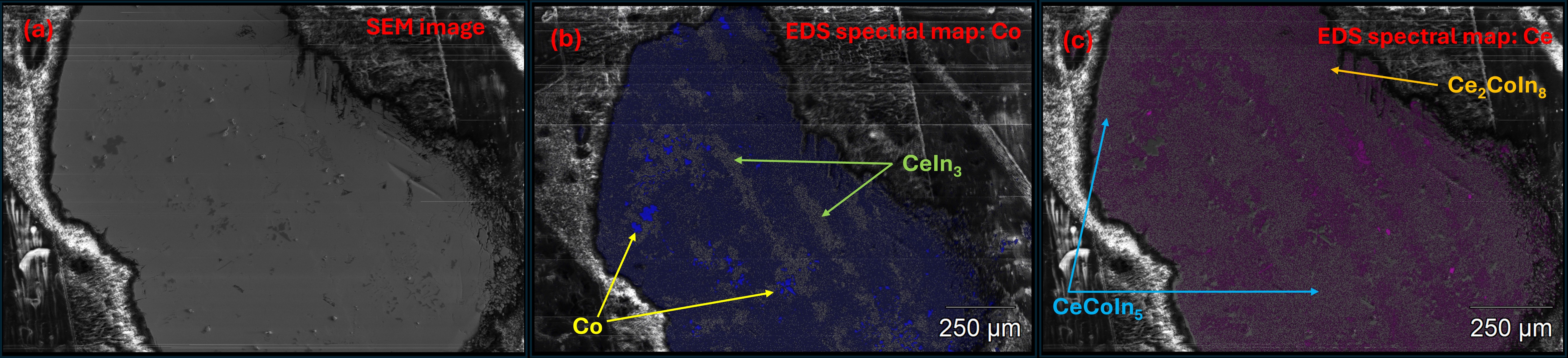}
\caption{\label{fig:EDSspectralmapshori} EDS spectral mapping of the polycrystalline CeCoIn$_5$ sample. (a) SEM images of the polycrystalline CeCoIn$_5$ sample. (b) Ce and (c) Co elemental composition maps overlaid on the SEM image. Few regions corresponding to different phases (CeCoIn$_5$,  Ce$_2$CoIn$_8$, CeIn$_3$ and free Co) are indicated by arrows.}
\end{figure*}

After EBSD imaging, we performed spectral mapping with energy-dispersive X-ray spectroscopy (EDS) to determine the elemental composition across the sample as shown in Fig. \ref{fig:EDSspectralmapshori}. The elemental composition maps for Co and Ce are overlaid on the SEM image of the sample in Fig. \ref{fig:EDSspectralmapshori}(b) and (c), respectively. From a correlative analysis between the EBSD maps and EDS spectral imaging we identified the different phases present in the sample. For example, the dark blue regions indicated in Fig. \ref{fig:EDSspectralmapshori}(b) correspond to free Co, while the Co deficient regions correspond to CeIn$_3$. Similarly, in Fig. \ref{fig:EDSspectralmapshori}(c) the regions corresponding to CeCoIn$_5$ and Ce$_2$CoIn$_8$ phases are identified based on the relative abundance of Ce and Co. The different phases (CeCoIn$_5$,  Ce$_2$CoIn$_8$, CeIn$_3$ and free Co) were further confirmed by performing EDS at several spots in each region determining the exact elemental composition.
\subsection{Details of EBSD setup shown in Fig. \ref{fig:EBSDtechplusIPF2}(a)}
A schematic of EBSD setup is shown in Fig. \ref{fig:EBSDtechplusIPF2}(a). In the SEM chamber, a beam of electrons is directed at the sample tilted at 70$^\circ$ from horizontal. A portion of the beam is backscattered from the sample forming Kikuchi diffraction pattern on the phosphor fluorescent screen of the EBSD detector. A CMOS-based detector images the diffraction pattern. The diffraction pattern is indexed against the crystal structure of the sample to obtain grains and grain boundary orientation information.
\subsection{Sample preparation for micromachining}
The thin polished sample was gently removed from the lapping fixture by immersing it in acetone to remove the crystal bond holding the sample. Once the crystal bond was completely dissolved and the sample appeared mobile in the acetone bath, it was gently transported to a sapphire substrate. The electrical contacts were then made by attaching 12.5 $\mu$m platinum wire to the sample using silver paint leading to contacts with resistance less than 10 $\Omega$. The silver paint at the contact also acts as the glue that holds the thin sample to the sapphire substrate, as shown in the SEM images in Figure \ref{fig:Transportdevicesmain}.
\begin{figure*}
	\centering
   \includegraphics[width=\linewidth]{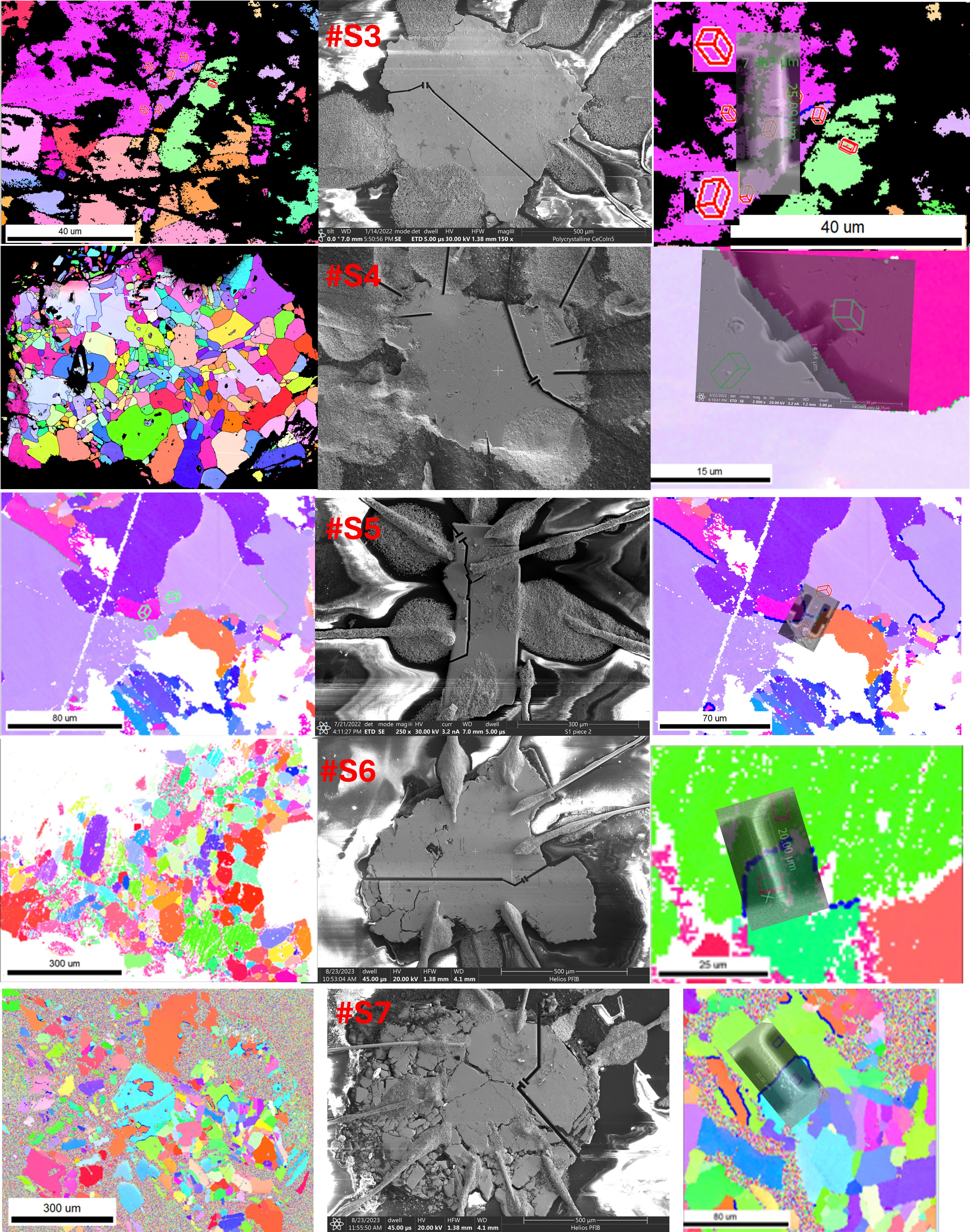}
\caption{\label{fig:Device1}Inverse pole figure (IPF) maps of polycrystaline CeCoIn$_5$ samples $S3$, $S5$, $S6$, and $S7$ depicting 90$^\circ$ misorientation grain boundary about $[100]$ and corresponding SEM images of the grain boundary devices. $S4$ device features a $70^\circ$ misorientation grain boundary about $[110]$.}
\end{figure*}
\begin{figure*}
	\centering
  \includegraphics[width=\linewidth]{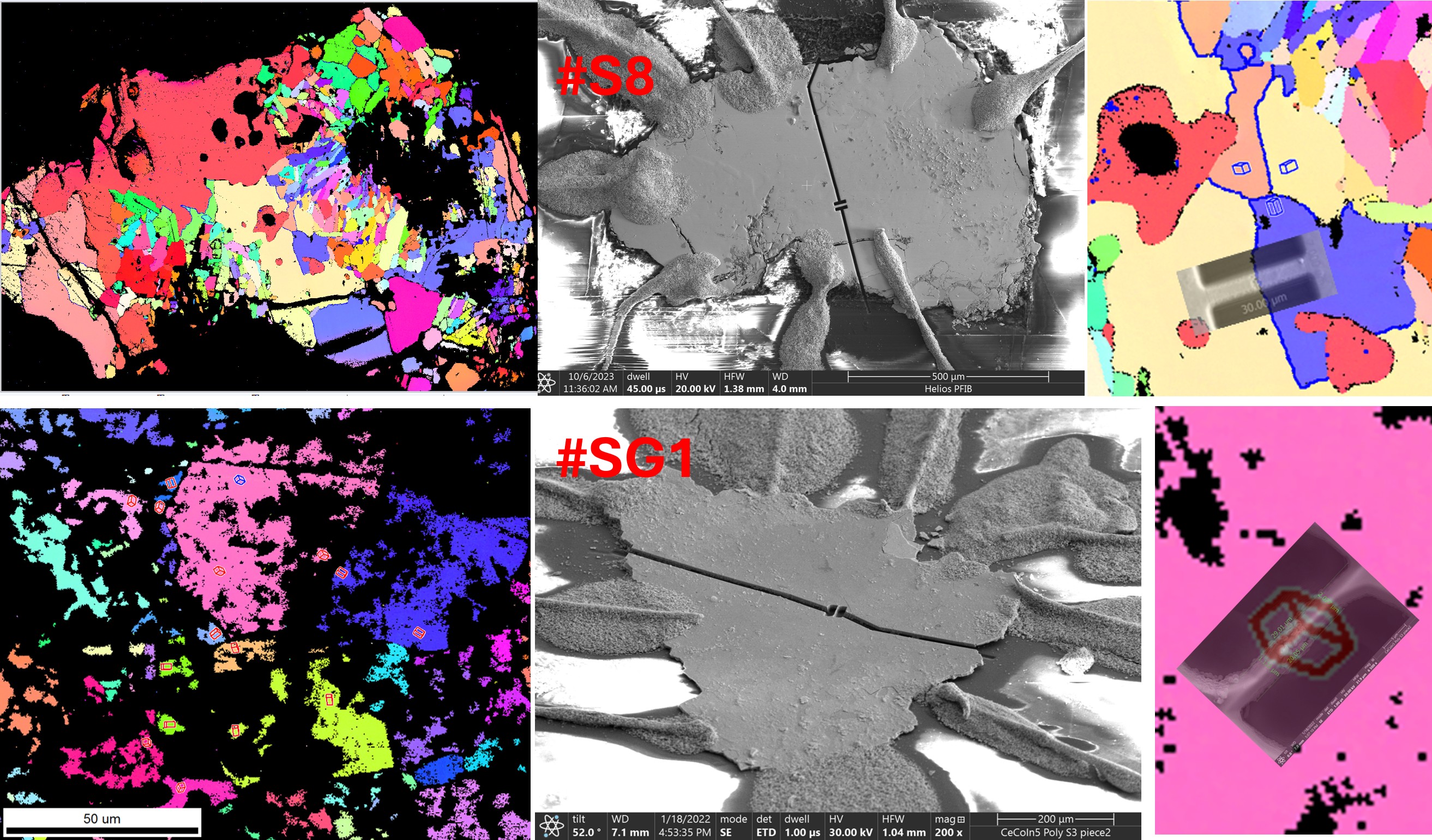}
\caption{\label{fig:Device2}Inverse pole figure (IPF) maps of polycrystaline CeCoIn$_5$ samples $S8$ and $SG1$. $S8$ depicts the 90$^\circ$ misorientation grain boundary about $[100]$ and corresponding SEM images of the grain boundary device. $SG1$ represents the single grain device showing only one drop in the superconducting transition in applied magnetic fields as discussed in the text.}
\end{figure*}
\subsection{Additional grain boundary devices}
Fig. \ref{fig:Device1} and Fig. \ref{fig:Device2} show the EBSD and SEM images of 90$^\circ$ grain boundary devices $S3$, $S5$, $S6$, $S7$, $S8$, the $70^\circ$ grain boundary device $S4$ and the single-grain device $SG1$ presented in the main text.


\begin{acknowledgments}
We acknowledge S-Z Lin for insightful discussions. Work at Los Alamos was supported by the U.S. Department of Energy, Office of Science, Basic Energy Sciences, Materials Sciences and Engineering Division, `Quantum fluctuations in narrow band systems' project. Scanning electron microscope, energy dispersive x-ray, electron backscatter diffraction imaging and focused ion-beam milling were performed at the Electron Microscopy Laboratory at Los Alamos National laboratory and the Center for Integrated Nanotechnologies, an Office of Science User Facility operated for the U.S. Department of Energy Office of Science. 
\end{acknowledgments}

\bibliography{CeCoIn5_microstructures_V3}
\end{document}